\documentclass[11pt]{article}

\usepackage{hyperref,amsfonts,amsmath,amssymb,graphicx,bm,subfigure,a4wide,cite}

\numberwithin{equation}{section}

\begin{document}

\title{\textbf{Thin layer axion dynamo}}

\author{Maxim Dvornikov\thanks{maxdvo@izmiran.ru}
\\
\small{\ Pushkov Institute of Terrestrial Magnetism, Ionosphere} \\
\small{and Radiowave Propagation (IZMIRAN),} \\
\small{108840 Moscow, Troitsk, Russia}}

\date{}

\maketitle

\begin{abstract}
We study interacting classical magnetic and pseudoscalar fields in
frames of the axion electrodynamics. A large scale pseudoscalar field
can be the coherent superposition of axions or axion like particles.
We consider the evolution of these fields in a thin spherical layer.
Decomposing the magnetic field into the poloidal and toroidal components,
we take into account their symmetry properties. The dependence of
the pseudoscalar field on the latitude is accounted for the induction
equation. Then, we derive the dynamo equations in the low mode approximation.
The nonlinear evolution equations for the harmonics of the magnetic
and pseudoscalar fields are solved numerically. As an application, we consider 
a dense axion star embedded in solar plasma. The behavior of the
harmonics and their typical oscillations frequencies are obtained.
We suggest that such small objects consisting of axions and confined
magnetic fields can cause the recently observed flashes in solar corona
contributing to its heating.
\end{abstract}

\section{Introduction}

The major fraction of the universe mass, called dark matter, almost
does not interact with light. Dark matter forms the halo throughout the
Galaxy, where it is distributed more or less uniformly. Nevertheless,
the presence of a measurable fraction of dark matter in the vicinity
of usual stars, like the Sun, is not excluded~\cite{Pet09}. Moreover,
a possible dark matter detection was recently reported in Ref.~\cite{Apr20}.
In principle, dark matter can form clusters and stars~\cite{BraZha19}
which are not related to baryonic astronomical objects. Thus, one
can consider spatially confined dark matter structures.

The origin and the content of dark matter is unclear. Axions and axion
like particles (ALP) are considered as the most plausible candidates
for dark matter~\cite{DufBib09}. Besides the gravitational interaction,
these particles interact rather weakly with electromagnetic fields~\cite{KimCar10}.
It can lead to numerous phenomena, like the emission of strong electromagnetic
radiation, in collisions of axion stars with, e.g., neutron stars
(see, e.g., Ref.~\cite{BraZha19}).

In frames of the axion magneto-hydrodynamics (MHD), a magnetic field
was mentioned in Ref.~\cite{LonVac15} to be unstable since the time
dependent axion wavefunction acts as the $\alpha$-dynamo parameter.
The axion MHD in the early universe was studied in Refs.~\cite{DvoSem20,HwaNoh22}.
The evolution of large scale magnetic fields in the presence of inhomogeneous
axions in the mean field approximation was considered in Ref.~\cite{Dvo22}.

The axion dynamo in neutron stars was developed in Ref.~\cite{Anz23}.
However, the induction equation used in Ref.~\cite{Anz23} does not
account for the coordinate dependence of the axion wavefunction. The
most complete induction equation, accounting for the axions spatial
inhomogeneity, was derived recently in Ref.~\cite{AkhDvo24} (see
also Appendix~\ref{sec:DERINDEQ}). Based on this equation, in Ref.~\cite{AkhDvo24},
we analyzed the mixing between two Chern-Simons waves in one dimensional
geometry, as well as a more sophisticated three dimensional (3D) case
involving the Hopf fibration. The emission of photons by axions in
strong magnetic fields in the vicinity of neutron stars was studied
in Ref.~\cite{DayMcd19}.

In the present work, based on the results of Ref.~\cite{AkhDvo24},
we develop a 3D axion dynamo in an axion spherical star. We start
in Sec.~\ref{sec:DYN} with deriving of the differential equations
for the poloidal and toroidal magnetic fields, as well as for the
axion wavefunction. Then, in Sec.~\ref{sec:APPL}, we consider the
application of our results for the evolution of magnetic fields in
an axion star embedded in solar plasma. Finally, we conclude in Sec.~\ref{sec:CONCL}.
Some phenomenological consequences of our results are also discussed
in Sec.~\ref{sec:CONCL}. The modified induction equation accounting
for the spatially inhomogeneous axion wavefunction is rederived in
Appendix~\ref{sec:DERINDEQ}. The system of nonlinear differential
equations for the harmonics of magnetic and pseudoscalar fields are
obtained in Appendix~\ref{sec:DIFFEQHARM}.

\section{Axion dynamo\label{sec:DYN}}

In this section, we derive the main dynamo equations in frames of
the axion MHD in the low mode approximation. We consider the dynamo
action in a thin spherical layer.

The evolution of the magnetic field $\mathbf{B}$ under the influence
of the external inhomogeneous pseudoscalar field $\varphi$ obeys
the equation (see Eq.~(\ref{eq:indeqgen}) and Ref.~\cite{AkhDvo24}),
\begin{equation}\label{eq:indeq}
  \dot{\mathbf{B}}=\nabla\times
  \left[
    \mathbf{b}\times(\nabla\times\mathbf{B})+\alpha\mathbf{B}-\eta(\nabla\times\mathbf{B})
  \right].
\end{equation}
If the pseudoscalar field is a coherent superposition of axions, $\alpha=g_{a\gamma}\eta\dot{\varphi}$
is the $\alpha$-dynamo parameter, $\mathbf{b}=g_{a\gamma}\eta^{2}\nabla\varphi$
is the axial vector accounting for the spatial inhomogeneity of $\varphi$,
$\eta$ is the magnetic diffusion coefficient, and $g_{a\gamma}$
is the axion-photon coupling constant. In Eq.~(\ref{eq:indeq}),
a dot means the time derivative. The molecular contribution to the
magnetic diffusion coefficient is $\eta=\sigma^{-1}$, where $\sigma$
is the electric conductivity. However, the turbulent magnetic diffusion
can be much sizable than the molecular one.

The induction Eq.~(\ref{eq:indeq}) should be supplied with the inhomogeneous
Klein-Gordon equation for $\varphi$~\cite{Dvo22,AkhDvo24},
\begin{equation}\label{eq:KGeq}
\ddot{\varphi}-\Delta\varphi+m^{2}\varphi=g_{a\gamma}(\mathbf{EB}),
\end{equation}
where $m$ is the mass of $\varphi$. The expression for the electric
field is
\begin{equation}\label{eq:E}
  \mathbf{E}=\eta(\nabla\times\mathbf{B})-\alpha\mathbf{B}-[\mathbf{b}\times(\nabla\times\mathbf{B})],
\end{equation}
which is given in Eq.~(\ref{eq:Egen}) (see also Refs.~\cite{Dvo22,AkhDvo24}).

We consider the fields $\mathbf{B}$ and $\varphi$ inside the spherical
volume which can be an axion star. All the quantities in Eqs.~(\ref{eq:indeq})
and~(\ref{eq:KGeq}) are supposed to be axially symmetric. For example,
$\alpha=\alpha(r,\theta,t)$, $\mathbf{b}=b_{r}\mathbf{e}_{r}+b_{\theta}\mathbf{e}_{\theta}$,
and $b_{r,\theta}=b_{r,\theta}(r,\theta,t)$. Here we use the orthonormal
basis in spherical coordinates $\mathbf{e}_{i}$, $i=r,\theta,\phi$.
The $\alpha$-dynamo parameter should by antisymmetric with respect
to the equatorial plane, $\alpha(r,\pi-\theta,t)=-\alpha(r,\theta,t)$,
since $\varphi$ is pseudoscalar. The magnetic field is separated
into the poloidal $\mathbf{B}_{p}$ and toroidal $\mathbf{B}_{t}$
components, $\mathbf{B}=\mathbf{B}_{p}+\mathbf{B}_{t}$. We take that
$\mathbf{B}_{p}=\nabla\times(A\mathbf{e}_{\phi})$ and $\mathbf{B}_{t}=B\mathbf{e}_{\phi}$.
The new functions $A$ and $B$ have the following symmetry properties:
$A(r,\pi-\theta,t)=A(r,\theta,t)$ and $B(r,\pi-\theta,t)=-B(r,\theta,t)$.

Making tedious but straightforward calculations based on Eq.~(\ref{eq:indeq}),
we get the equations for $A$ and $B$,
\begin{align}\label{eq:ABeqs}
  \frac{\partial A}{\partial t}= & -g_{a\gamma}\eta^{2}\frac{1}{r}
  \left[
    \frac{\partial\varphi}{\partial r}\frac{\partial}{\partial r}
    \left(
      rB
    \right)+
    \frac{1}{r\sin\theta}\frac{\partial\varphi}{\partial\theta}\frac{\partial}{\partial\theta}
    \left(
      \sin\vartheta B
    \right)
  \right] +
  g_{a\gamma}\eta B\frac{\partial\varphi}{\partial t}+\eta\Delta'A,
  \nonumber
  \\
  \frac{\partial B}{\partial t}= & g_{a\gamma}\eta^{2}\frac{1}{r}
  \left[
    \frac{\partial}{\partial r}
    \left(
      r\frac{\partial\varphi}{\partial r}\Delta'A
    \right) +
    \frac{1}{r}\frac{\partial}{\partial\theta}
    \left(
      \frac{\partial\varphi}{\partial\theta}\Delta'A
    \right)
  \right]
  \nonumber
  \\
  & -
  g_{a\gamma}\eta\frac{1}{r}
  \left[
    \frac{\partial}{\partial r}
    \left(
      \frac{\partial\varphi}{\partial t}\frac{\partial}{\partial r}
      \left(
        rA
      \right)
    \right) +
    \frac{1}{r}\frac{\partial}{\partial\theta}
    \left(
      \frac{1}{\sin\theta}\frac{\partial\varphi}{\partial t}\frac{\partial}{\partial\theta}
      \left(
        \sin\theta A
      \right)
    \right)
  \right] +
  \eta\Delta'B,
\end{align}
where $\Delta'=\Delta-\tfrac{1}{r^{2}\sin^{2}\theta}=\tfrac{1}{r^{2}}\tfrac{\partial}{\partial r}\left(r^{2}\tfrac{\partial}{\partial r}\right)+\tfrac{1}{r^{2}\sin\theta}\tfrac{\partial}{\partial\theta}\left(\sin\theta\tfrac{\partial}{\partial\theta}\right)-\tfrac{1}{r^{2}\sin^{2}\theta}$
is the modified Laplace operator. Analogously, we transform Eq.~(\ref{eq:KGeq})
to the form,
\begin{multline}\label{eq:phieq}
  \frac{\partial^{2}\varphi}{\partial t^{2}}+m^{2}\varphi-\frac{1}{r^{2}}\frac{\partial\varphi}{\partial r}
  \left(
    r^{2}\frac{\partial\varphi}{\partial r}
  \right) -
  \frac{1}{r^{2}\sin\theta}\frac{\partial}{\partial\theta}
  \left(
    \sin\theta\frac{\partial\varphi}{\partial\theta}
  \right)
  \\
  = g_{a\gamma}\eta
  \left[
    \frac{1}{r^{2}\sin^{2}\theta}\frac{\partial}{\partial\theta}
    \left(
      \sin\vartheta B
    \right)\frac{\partial}{\partial\theta}
    \left(
      \sin\vartheta A
    \right) +
    \frac{1}{r^{2}}\frac{\partial}{\partial r}
    \left(
      rB
    \right)
    \frac{\partial}{\partial r}
    \left(
      rA
    \right) -
    B\Delta'A
  \right].
\end{multline}
To derive Eq.~(\ref{eq:phieq}) we keep only the first term in the
right hand side of Eq.~(\ref{eq:E}) to guarantee that the result
is linear in $g_{a\gamma}$.

Now, following Ref.~\cite{Par55}, we assume that the fields $(A,B,\varphi)$
evolve in a thin layer between $R$ and $R+\mathrm{d}r$, where $R$
is the typical size of an axion star and $\mathrm{d}r\ll R$. In this
case, the radial dependence of the functions can be neglected. Therefore,
we can replace $r\to R$ and $\frac{\partial}{\partial r}\to0$ in
Eqs.~(\ref{eq:ABeqs}) and~(\ref{eq:phieq}).

Using the dimensionless variables
\begin{equation}\label{eq:dmnlsvar}
  \mathcal{A}=g_{a\gamma}A,
  \quad
  \mathcal{B}=g_{a\gamma}RB,
  \quad
  \Phi=\frac{g_{a\gamma}\eta}{R}\varphi,
  \quad
  \tau=\frac{\eta t}{R^{2}},
\end{equation}
we rewrite Eqs.~(\ref{eq:ABeqs}) and~(\ref{eq:phieq}) in the form,
\begin{align}\label{eq:ABPhieq}
  \frac{\partial\mathcal{A}}{\partial\tau}= & -\frac{\partial\Phi}{\partial\theta}
  \left[
    \frac{\partial\mathcal{B}}{\partial\theta}+\cot\theta\mathcal{B}
  \right]+
  \frac{\partial\Phi}{\partial\tau}\mathcal{B}+\frac{\partial^{2}\mathcal{A}}{\partial\theta^{2}}+
  \cot\theta\frac{\partial\mathcal{A}}{\partial\theta}-\frac{\mathcal{A}}{\sin^{2}\theta},
  \nonumber
  \\
  \frac{\partial\mathcal{B}}{\partial\tau}= & \frac{\partial^{2}\Phi}{\partial\theta^{2}}
  \left(
    \frac{\partial^{2}\mathcal{A}}{\partial\theta^{2}}+\cot\theta\frac{\partial\mathcal{A}}{\partial\theta}-\frac{\mathcal{A}}{\sin^{2}\theta}
  \right)
  \nonumber
  \\
  & +
  \frac{\partial\Phi}{\partial\theta}
  \left(
    \frac{\partial^{3}\mathcal{A}}{\partial\theta^{3}}+\cot\theta\frac{\partial^{2}\mathcal{A}}{\partial\theta^{2}}-
    \frac{2}{\sin^{2}\theta}\frac{\partial\mathcal{A}}{\partial\theta}+\frac{2\cot\theta}{\sin^{2}\theta}\mathcal{A}
  \right)
  \nonumber
  \\
  & -
  \frac{\partial^{2}\Phi}{\partial\tau\partial\theta}
  \left(
    \frac{\partial\mathcal{A}}{\partial\theta}+\cot\theta\mathcal{A}
  \right)-
  \frac{\partial\Phi}{\partial\tau}
  \left(
    \frac{\partial^{2}\mathcal{A}}{\partial\theta^{2}}+\cot\theta\frac{\partial\mathcal{A}}{\partial\theta}-
    \frac{\mathcal{A}}{\sin^{2}\theta}
  \right)
  \nonumber
  \\
  & +
  \frac{\partial^{2}\mathcal{B}}{\partial\theta^{2}}+\cot\theta\frac{\partial\mathcal{B}}{\partial\theta}-\frac{\mathcal{B}}{\sin^{2}\theta},
  \nonumber
  \displaybreak[2]
  \\
  \frac{\partial^{2}\Phi}{\partial\tau^{2}}= & -\mu^{2}\Phi+\kappa^{2}
  \left(
    \frac{\partial^{2}\Phi}{\partial\theta^{2}}+\cot\theta\frac{\partial\Phi}{\partial\theta}
  \right)
  \nonumber
  \\
  & +
  \left(
    \frac{\partial\mathcal{B}}{\partial\theta}+\cot\theta\mathcal{B}
  \right)
  \left(
    \frac{\partial\mathcal{A}}{\partial\theta}+\cot\theta\mathcal{A}
  \right)+
  \mathcal{A}\mathcal{B}
  \nonumber
  \\
  & -
  \mathcal{B}
  \left(
    \frac{\partial^{2}\mathcal{A}}{\partial\theta^{2}}+\cot\theta\frac{\partial\mathcal{A}}{\partial\theta}-\frac{\mathcal{A}}{\sin^{2}\theta}
  \right),
\end{align}
where $\mu=mR^{2}/\eta$ is the dimensionless axion mass and $\kappa=R/\eta$
is the effective wave vector.

According to Ref.~\cite{NefSok10}, we decompose the dimensionless
functions $(\mathcal{A},\mathcal{B},\Phi)$ into the harmonics,
\begin{align}\label{eq:ABPhiharm}
  \mathcal{A} & =a_{1}(\tau)\sin\theta+a_{2}(\tau)\sin3\theta+\dotsc,
  \nonumber
  \\
  \mathcal{B} & =b_{1}(\tau)\sin2\theta+b_{2}(\tau)\sin4\theta+\dotsc,
  \nonumber
  \\
  \Phi & =\phi_{1}(\tau)\sin2\theta+\phi_{2}(\tau)\sin4\theta+\dotsc,
\end{align}
where the coefficients $a_{1,2}$, $b_{1,2}$, and $\phi_{1,2}$ are
the functions of $\tau$ only. The decomposition in Eq.~(\ref{eq:ABPhiharm})
obeys the symmetry conditions specified earlier. Note that use the
low mode approximation in Eq.~(\ref{eq:ABPhiharm}) considering only
two first harmonics. Substituting Eq.~(\ref{eq:ABPhiharm}) to Eq.~(\ref{eq:ABPhieq}),
we get the system of nonlinear ordinary differential equations, which
is provided in Eq.~(\ref{eq:abpsieq}), for the functions $a_{1,2}$,
$b_{1,2}$, and $\phi_{1,2}$.

\section{Axion dynamo in solar plasma\label{sec:APPL}}

Our main goal is to study the influence of an external pseudoscalar
field on the evolution of magnetic fields. For this purpose to assume
the existence of a spherical object consisting of coherent axions
where a seed magnetic field is present. An axion star is an example
of such a structure. We study the case of a dense axion star. Such
a star was found in Ref.~\cite{Bra16} be stable if its radius $R\sim(10^{-11}-10^{-10})R_{\odot}=(0.7-7)\,\text{cm}$
or $R\gtrsim10^{-4}R_{\odot}=70\,\text{km}$. The energy density of
axions in a dense axion star is $\rho\lesssim m^{2}f_{a}^{2}$~\cite{BraZha19},
where $f_{a}=\tfrac{\alpha_{\mathrm{em}}}{2\pi g_{a\gamma}}$ is the
Peccei--Quinn constant and $\alpha_{\mathrm{em}}=7.3\times10^{-3}$
is the fine structure constant.

We study the contribution of axions to the dynamics of magnetic fields
which can be present in solar plasma. We take the small radius of
an axion star $R=0.7\,\text{cm}$~\cite{Bra16}. The solar magnetic
diffusion coefficient was mentioned in Ref.~\cite[p.~370]{Sti04}
to be mainly turbulent one. We take that $\eta=10^{10}\,\text{cm}^{2}\cdot\text{s}^{-1}$,
which is close to the observed value given in Ref.~\cite{ChaLitSak08}.
The effective mass and the wave number are $\mu=7.5\times10^{-1}$
and $\kappa=2.1$. Here, we take that $m=10^{-5}\,\text{eV}$.

We suppose that $\dot{\varphi}(t=0)=0$ and the energy density of
axions is $\rho=10^{-2}m^{2}f_{a}^{2}$. In this case, the initial
value of $\rho$ is
\begin{equation}\label{eq:rho0}
  \rho_{0}=\frac{1}{2}
  \left[
    (\nabla\varphi_{0})^{2}+m^{2}\varphi_{0}^{2}
  \right] \approx
  \frac{\mu^{2}
  \left\langle
    (\partial_{\theta}\Phi_{0})^{2}
  \right\rangle
  }{2g_{a\gamma}^{2}R^{2}},
\end{equation}
where we use Eq.~(\ref{eq:dmnlsvar}) and the fact that $\mu^{2}\ll\kappa^{2}$.
The fact that $|\nabla\varphi_{0}|$ term is dominant in Eq.~(\ref{eq:rho0})
shows the importance the axion inhomogeneity in the system. Supposing
that $\phi_{2}(0)=0$, as well as taking that $\rho_{0}\approx10^{-2}m^{2}f_{a}^{2}$
and $\left\langle \cos2\theta\right\rangle =\tfrac{1}{2}$ in Eq.~(\ref{eq:rho0}),
we obtain the part of the initial condition for the system in Eq.~(\ref{eq:abpsieq}),
\begin{equation}
  \phi_{1}(0)=10^{-1}\frac{\alpha_{\mathrm{em}}mR}{2\pi\kappa}=2\times10^{-2},
\end{equation}
and $\phi_{2}(0)=0$, $\dot{\phi}_{1,2}(0)=0$.

The initial condition for $\mathcal{A}$ and $\mathcal{B}$ can be
obtained using Eq.~(\ref{eq:dmnlsvar}),
\begin{equation}\label{eq:a10b10}
  a_{1}(0)=1.4\times10^{-17}
  \left(
    \frac{B_{\mathrm{pol}}^{(0)}}{\mathrm{kG}}
  \right),
  \quad
  b_{1}(0)=1.4\times10^{-17}
  \left(
    \frac{B_{\mathrm{tor}}^{(0)}}{\mathrm{kG}}
  \right),
\end{equation}
and $a_{2}(0)=b_{2}(0)=0$. In Eq.~(\ref{eq:a10b10}), $B_{\mathrm{pol,tor}}^{(0)}$
are the seed poloidal and toroidal magnetic fields. We take that $B_{\mathrm{pol,tor}}^{(0)}=4\,\text{kG}$~\cite{Liv06}.
After setting the initial condition, we can solve the system in Eq.~(\ref{eq:abpsieq}).

\begin{figure}
  \centering
  \subfigure[]
  {\label{fig:fig1a}
  \includegraphics[scale=.35]{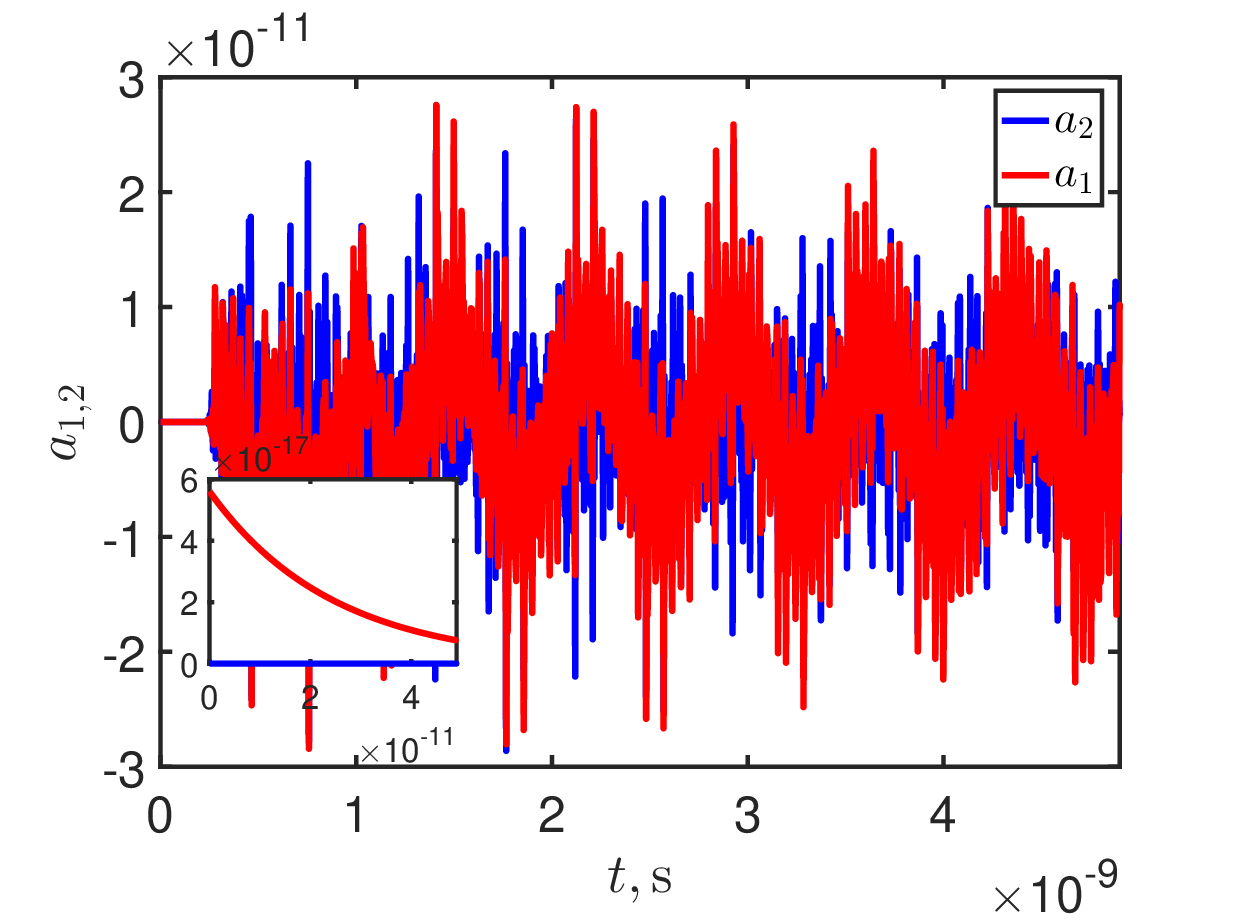}}
  \subfigure[]
  {\label{fig:fig1b}
  \includegraphics[scale=.35]{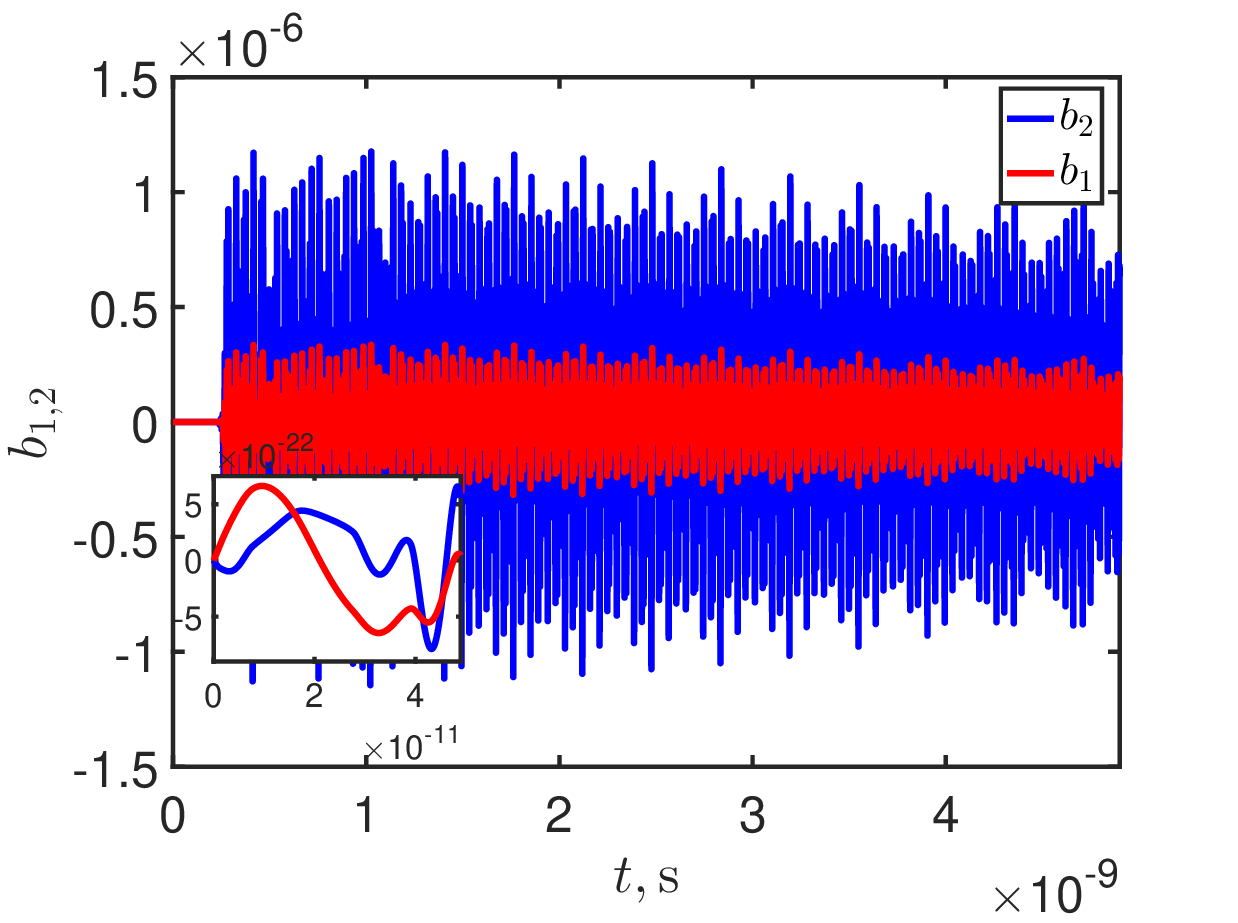}}
  \\
  \subfigure[]
  {\label{fig:fig1c}
  \includegraphics[scale=.35]{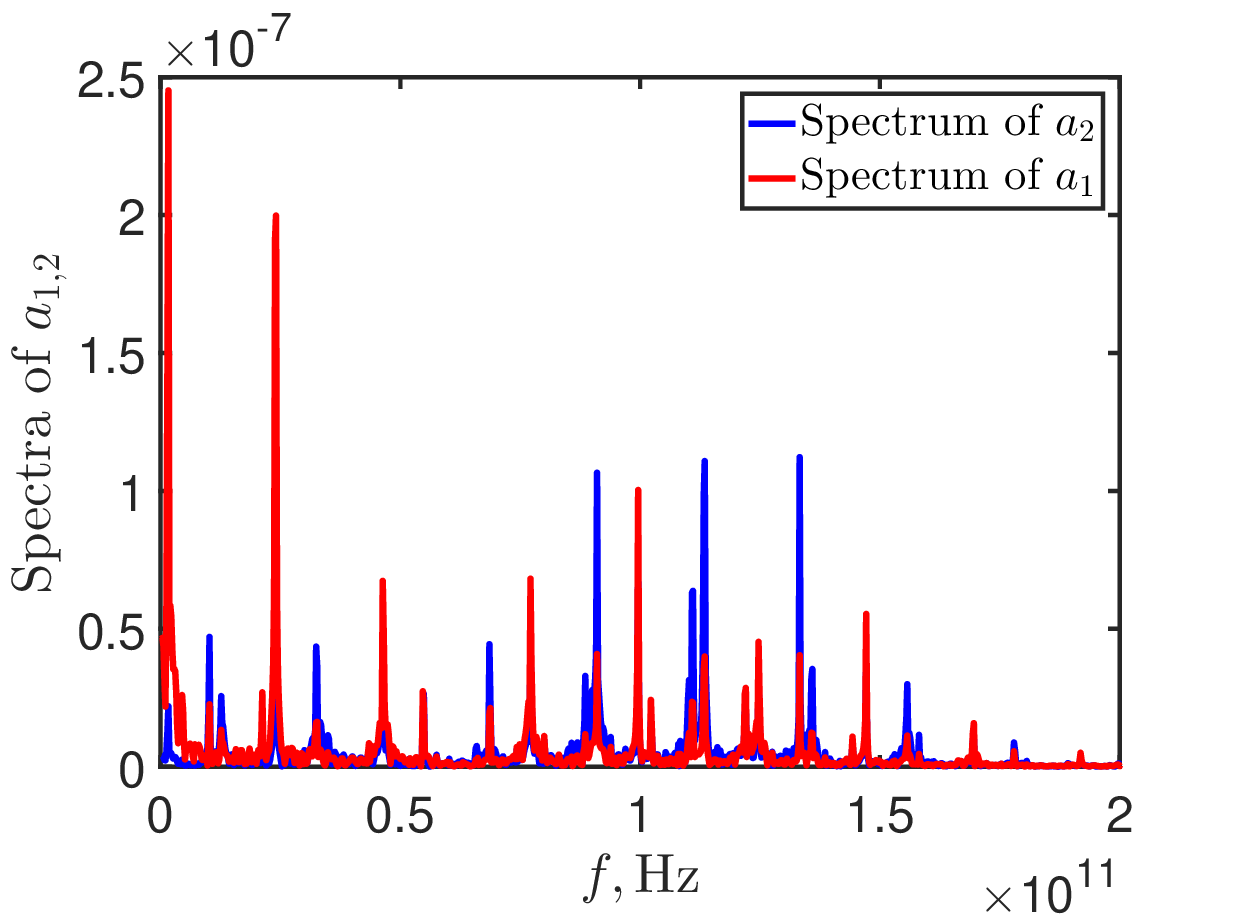}}
  \subfigure[]
  {\label{fig:fig1d}
  \includegraphics[scale=.35]{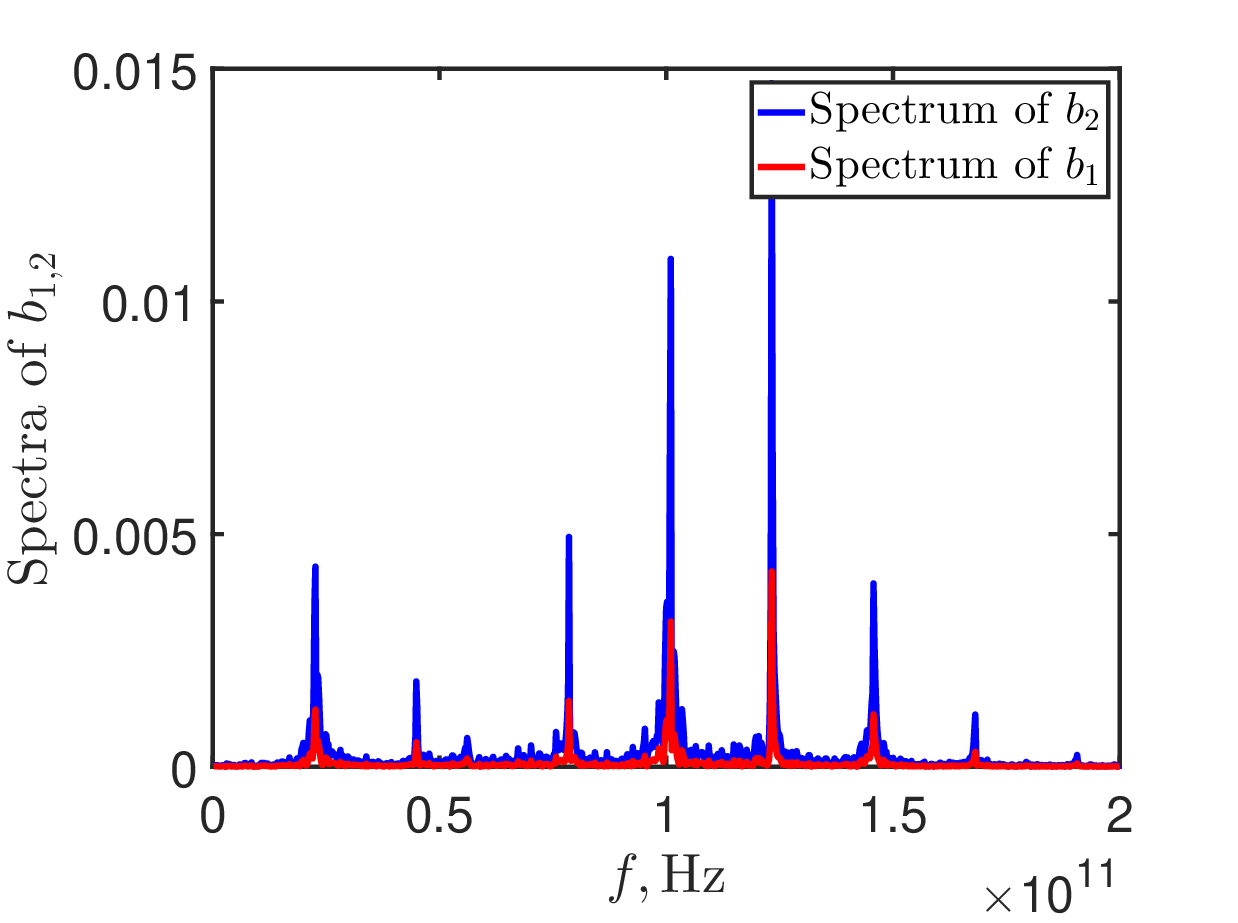}}
  \\
  \subfigure[]
  {\label{fig:fig1e}
  \includegraphics[scale=.35]{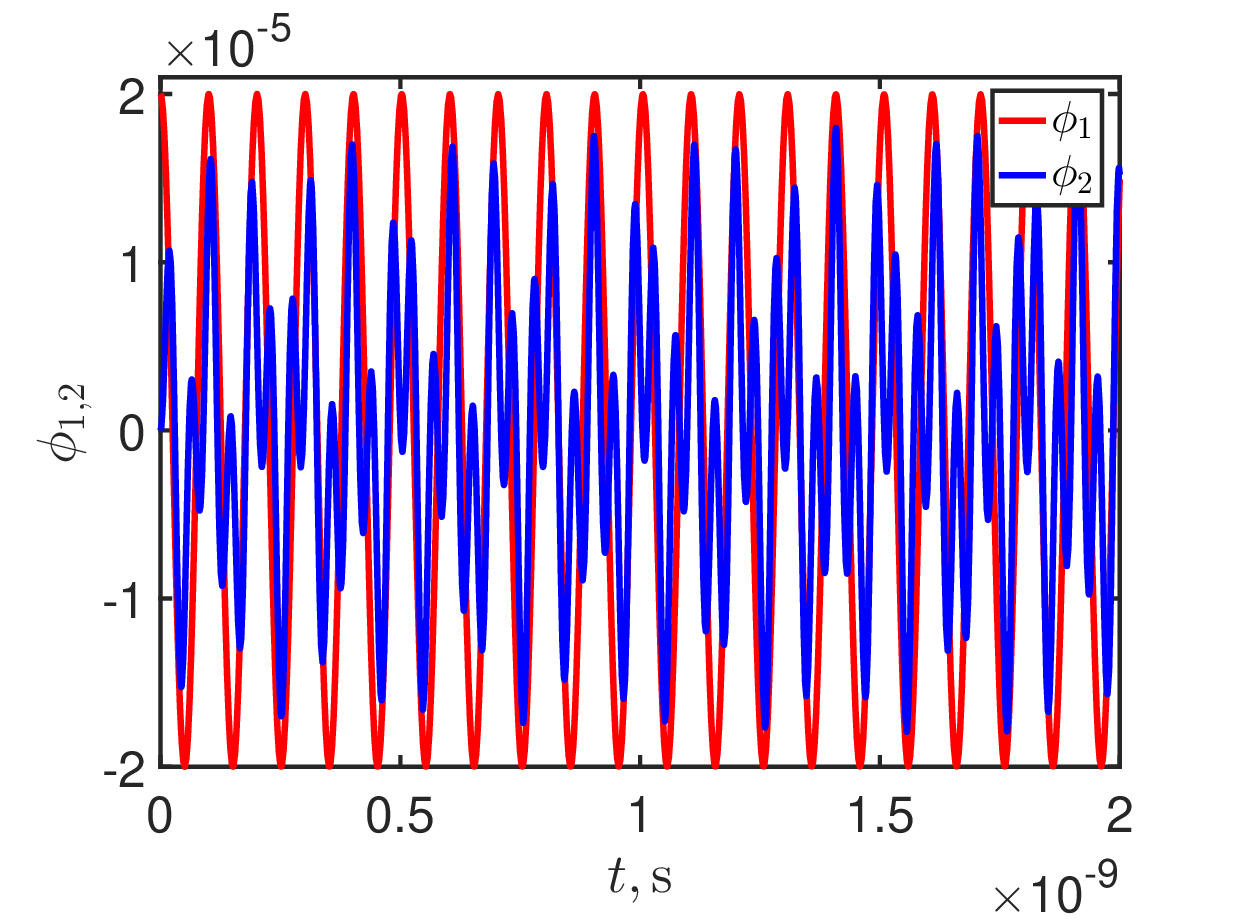}}
  \subfigure[]
  {\label{fig:fig1f}
  \includegraphics[scale=.35]{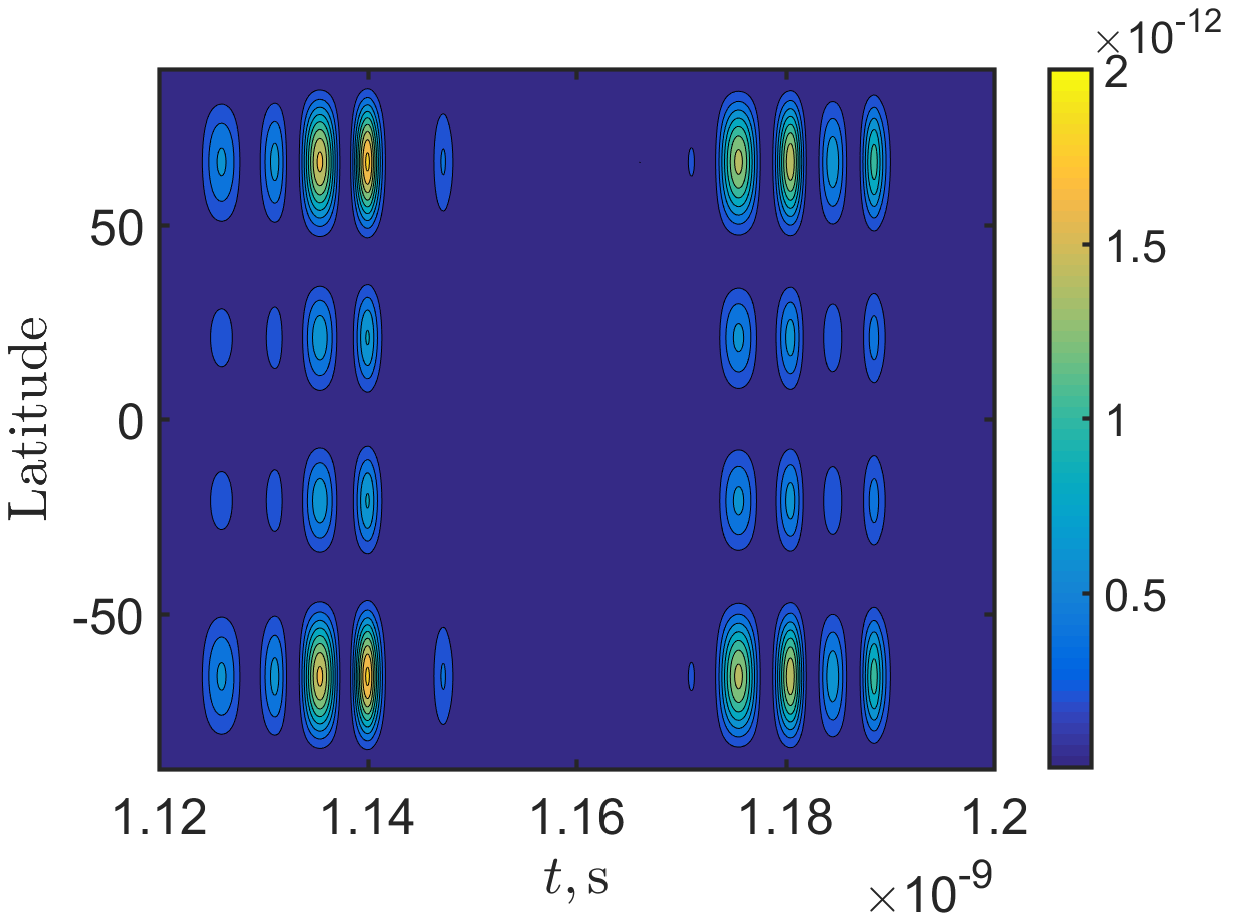}}
  \protect 
\caption{The evolution of the magnetic and pseudoscalar fields inside the axion
star embedded in solar plasma. (a) The time evolution of the poloidal
harmonics $a_{1,2}$. (b) The behavior of toroidal harmonics $b_{1,2}$
versus time. (c) The spectra of $a_{1,2}$. (d) The spectra of $b_{1,2}$.
(e) The harmonics $\phi_{1,2}$ of the pseudoscalar field. (f) The
total magnetic energy density in Eq.~(\ref{eq:magnen}). We take
that $B_{\mathrm{pol}}^{(0)}=4\,\text{kG}$ and $B_{\mathrm{tor}}^{(0)}=0$.\label{fig:sunpol}}
\end{figure}

In Fig.~\ref{fig:sunpol}, we show the behavior of the system in
solar plasma when only the seed poloidal magnetic field is present,
i.e. $a_{1}(0)\neq0$ and $b_{1}(0)=0$ in Eq.~(\ref{eq:a10b10}).
One can see in Figs.~\ref{fig:fig1a} and~\ref{fig:fig1b}
the evolution of the harmonics $a_{1,2}$ and $b_{1,2}$. The insets
in Figs.~\ref{fig:fig1a} and~\ref{fig:fig1b} represent
the behavior of these functions in small evolution times, when the
initial condition is visible.

The spectra of $a_{1,2}$ and $b_{1,2}$ are shown in Figs.~\ref{fig:fig1c}
and~\ref{fig:fig1d}. The poloidal component can be measured
since it extends to outer regions of an axion star. The typical frequency
of $a_{1}$ oscillations, in Fig.~\ref{fig:fig1c}, is $f\sim10^{10}\,\text{Hz}$.
Here we take the second peak in the spectrum of $a_{1}$. Such oscillations
frequency implies the validity of the causality condition, $fR<1$.
Moreover, the MHD approximation, $\eta^{-1}\gg f$, is also valid
in this case.

We depict the evolution of the total magnetic energy density in Fig.~\ref{fig:fig1e},
\begin{equation}\label{eq:magnen}
  \rho_{\mathrm{B}}(\theta,t)=\frac{\mathbf{B}^{2}}{2} \propto
  \left(
    \frac{\partial\mathcal{A}}{\partial\theta}+\cot\theta\mathcal{A}
  \right)^{2} +
  \mathcal{A}^{2}+\mathcal{B}^{2},
\end{equation}
in a short time interval to demonstrate its distribution over the
$\text{Latitude}=90^{\circ}\times\left(1-2\theta/\pi\right)$. It
is the analogue of a `butterfly' diagram in solar physics (see, e.g.,
Ref.~\cite[p.~377]{Sti04}).

The evolution of the harmonics of the pseudoscalar field is present
in Fig.~\ref{fig:fig1f}. Both harmonics of $\varphi$ have approximately
equal amplitudes. It demonstrates the importance of keeping the coordinate
dependence of $\varphi$ both in Eqs.~(\ref{eq:indeq}) and~(\ref{eq:KGeq}).
One can see that frequencies of $\varphi$ oscillations are much smaller
than those of the magnetic fields. 

\begin{figure}
  \centering
  \subfigure[]
  {\label{fig:fig2a}
  \includegraphics[scale=.35]{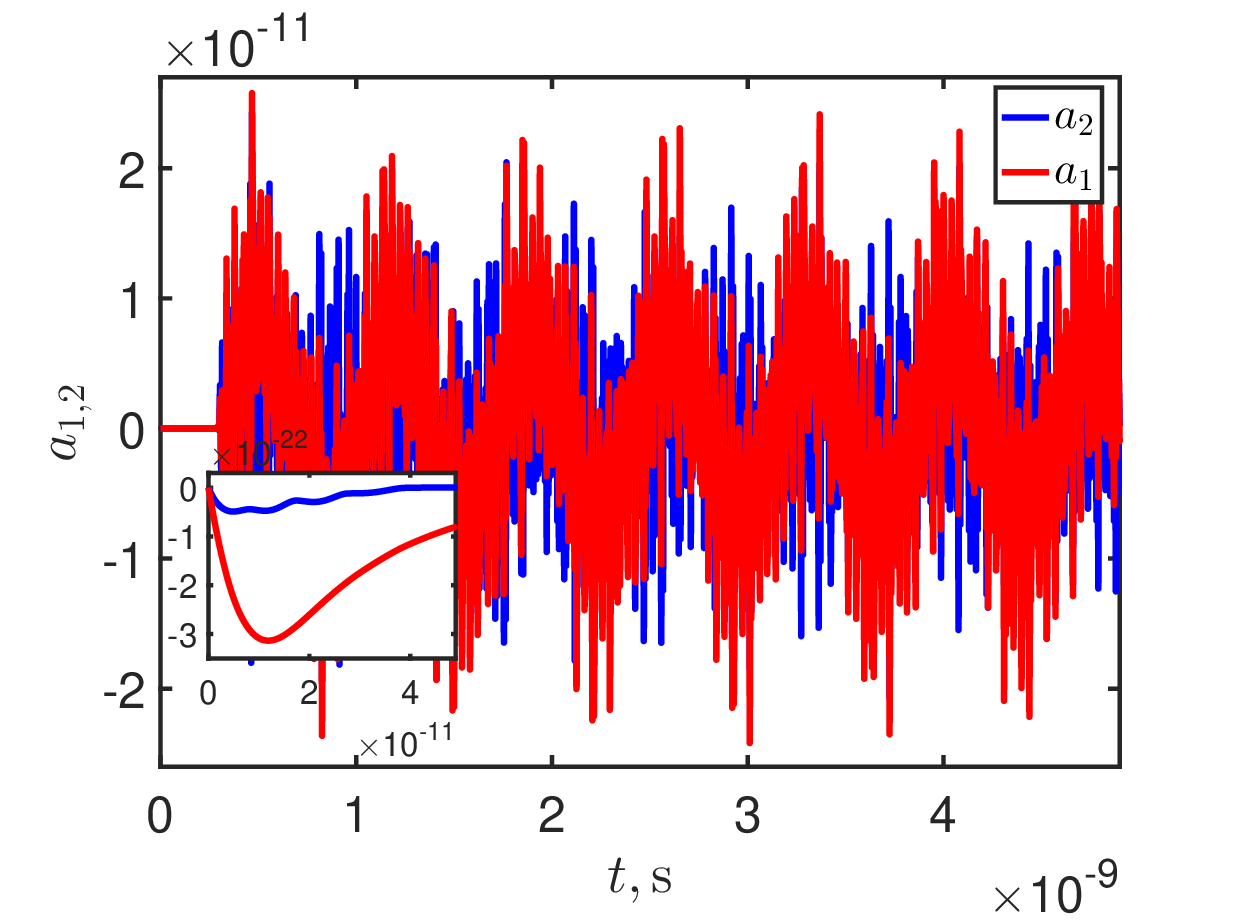}}
  \subfigure[]
  {\label{fig:fig2b}
  \includegraphics[scale=.35]{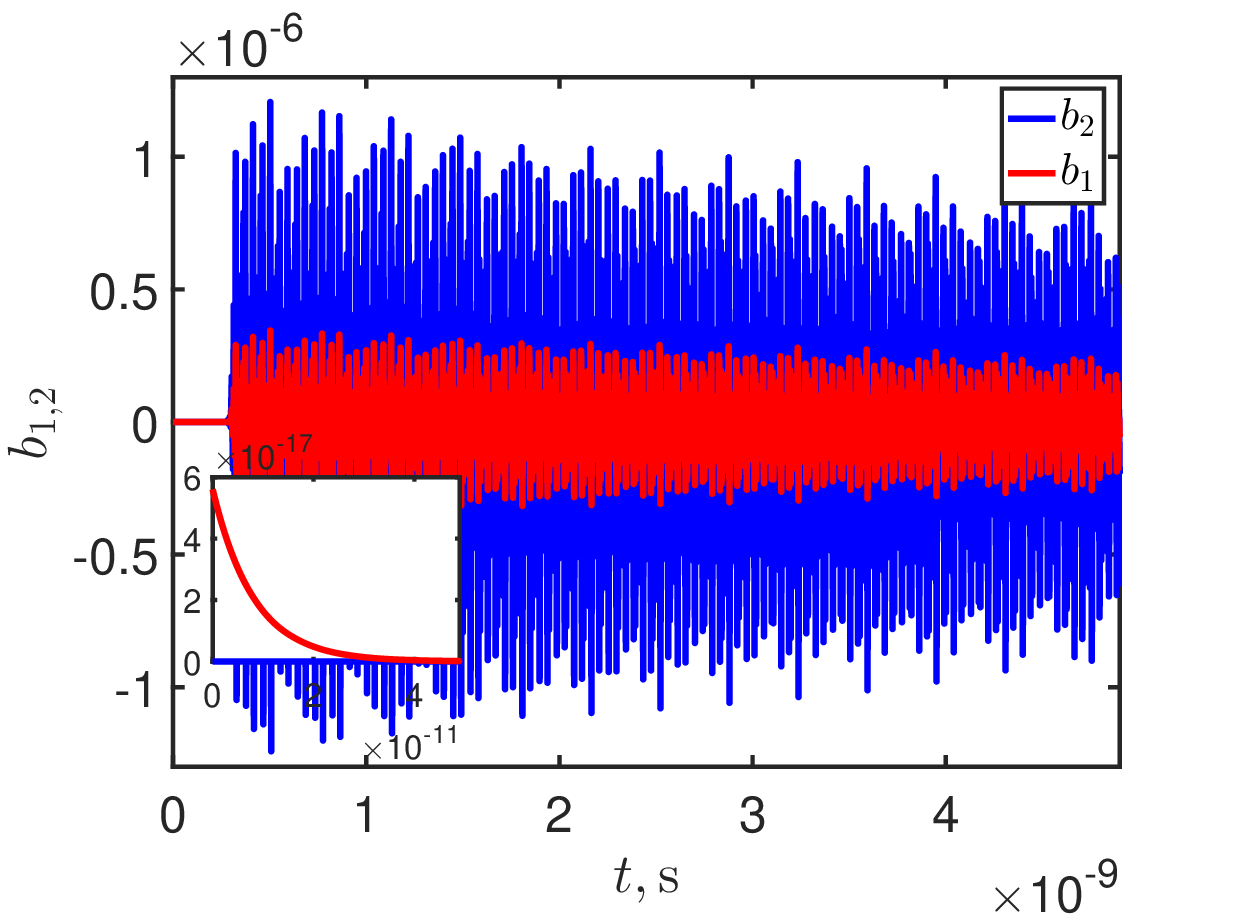}}
  \\
  \subfigure[]
  {\label{fig:fig2c}
  \includegraphics[scale=.35]{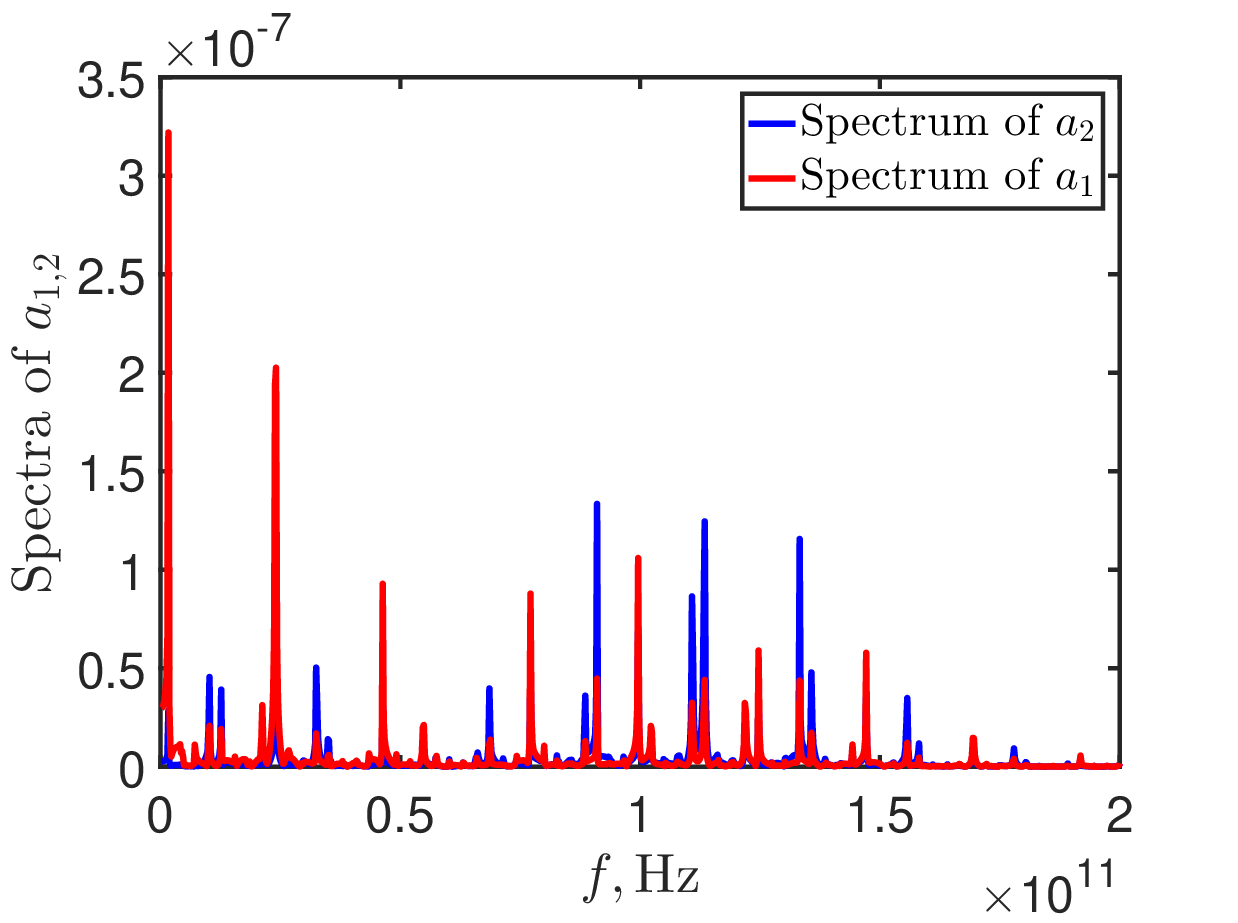}}
  \subfigure[]
  {\label{fig:fig2d}
  \includegraphics[scale=.35]{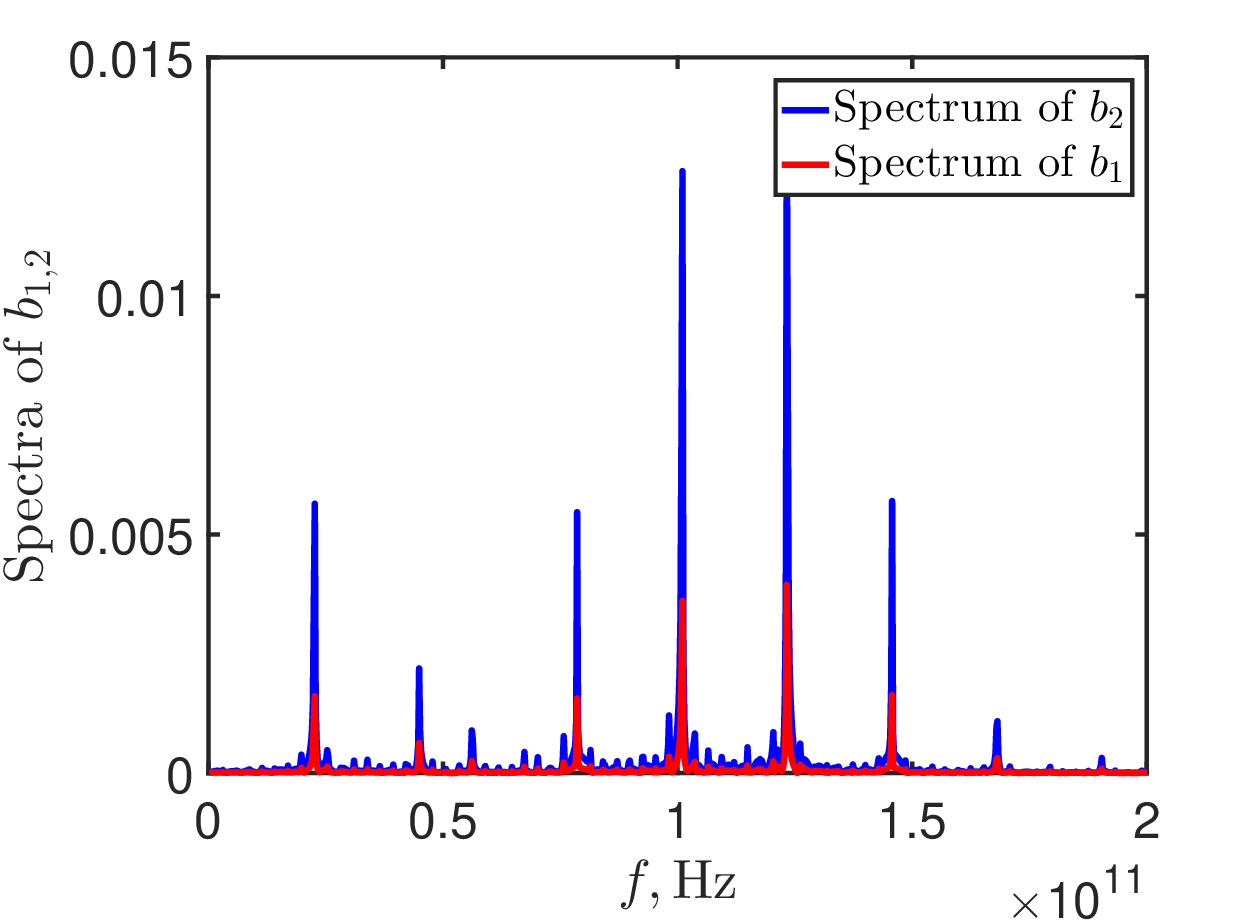}}
  \\
  \subfigure[]
  {\label{fig:fig2e}
  \includegraphics[scale=.35]{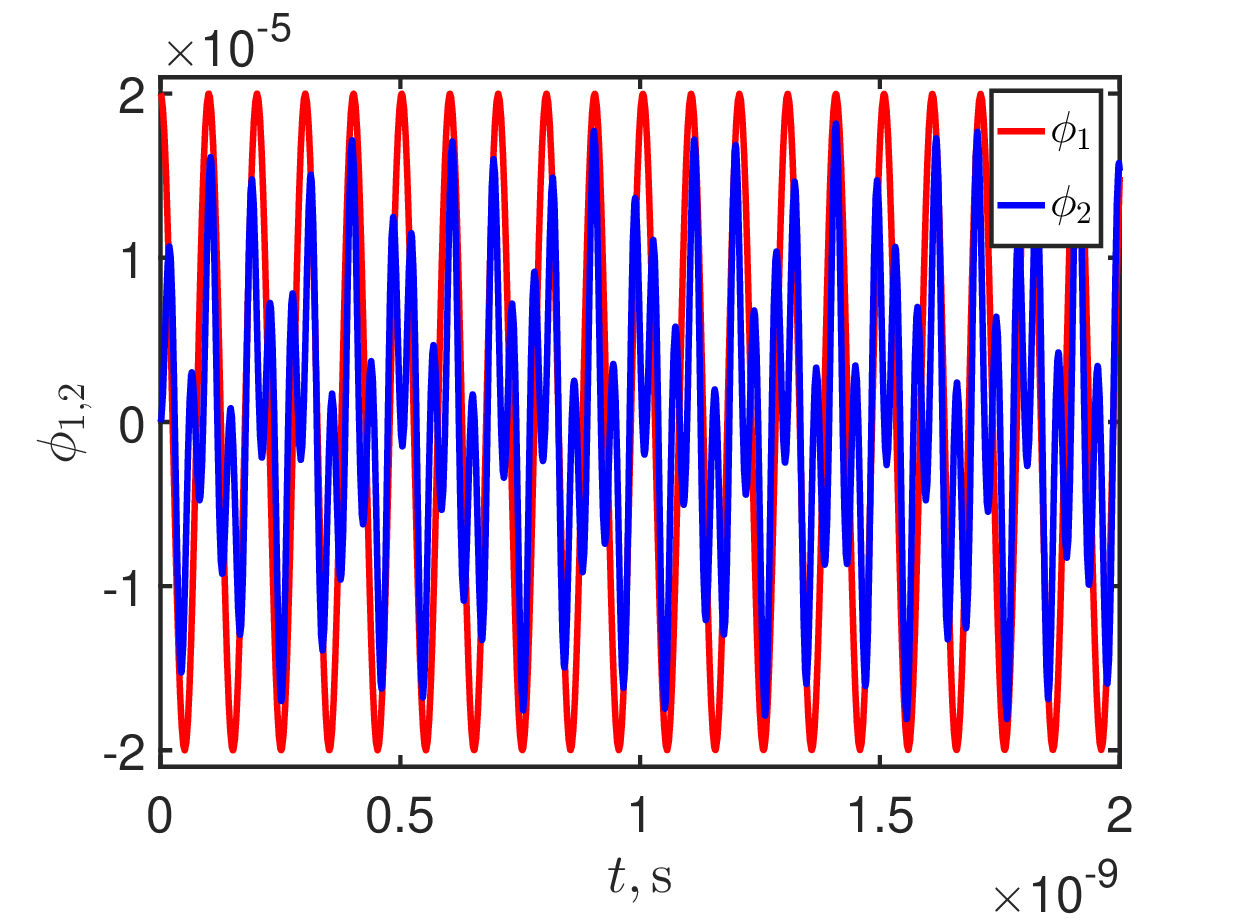}}
  \subfigure[]
  {\label{fig:fig2f}
  \includegraphics[scale=.35]{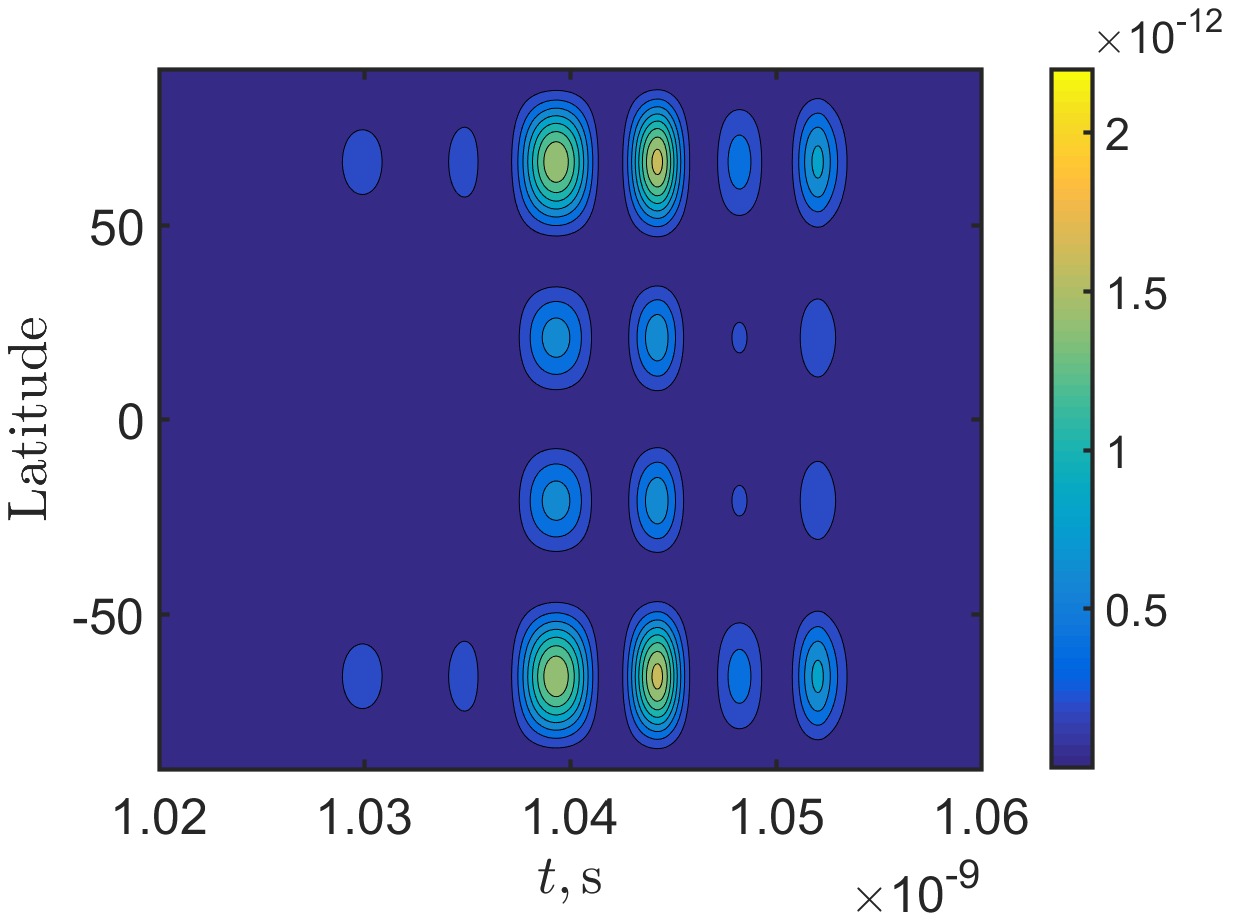}}
  \protect 
\caption{The same as in Fig.~\ref{fig:sunpol} for $B_{\mathrm{pol}}^{(0)}=0$
and $B_{\mathrm{tor}}^{(0)}=4\,\text{kG}$.\label{fig:suntor}}
\end{figure}

In Fig.~\ref{fig:suntor}, we show the evolution of the system
when only a seed toroidal magnetic field is present initially, i.e.
we take that $b_{1}(0)\neq0$ and $a_{1}(0)=0$. Now, the strength
of the seed toroidal field is the same as for the poloidal one in
Fig.~\ref{fig:sunpol}, i.e. $B_{\mathrm{tor}}^{(0)}=4\,\text{kG}$.
The initial condition for the harmonics can be seen in the insets
in Figs.~\ref{fig:fig2a} and~\ref{fig:fig2b}. The behavior
of the magnetic fields and $\varphi$ qualitatively resembles that
in Fig.~\ref{fig:sunpol}.

We can see in Figs.~\ref{fig:fig1f} and~\ref{fig:fig2f}
that the evolution of the pseudoscalar field is unaffected by the
magnetic field. We also notice in Figs.~\ref{fig:fig1a}, \ref{fig:fig1b},
\ref{fig:fig2a}, and~\ref{fig:fig2b} that the magnetic
fields are amplified by the $\alpha$-dynamo driven by the inhomogeneous
axion $\varphi$. After the amplification, the magnetic field enters
to the oscillating regime. It should be mentioned that, in Figs.~\ref{fig:fig1c}
and~\ref{fig:fig2c}, the frequency of the first peaks in
the spectra of $a_{1}$, which are the maximal ones, is $\sim10^{9}\,\text{Hz}$.

\section{Conclusion\label{sec:CONCL}}

In the present work, we have studied the simultaneous evolution of
the magnetic and pseudoscalar macroscopic fields. The latter can be
a coherent superposition of axions or ALP. This system obeys the axion
electrodynamics equations which result from the Lagrangian in Eq.~(\ref{eq:Lagr}).

In Sec.~\ref{sec:DYN}, based on the axion electrodynamics Eqs.~(\ref{eq:Maxeq1})-(\ref{eq:Maxeq4}),
we have derived the modified induction Eq.~(\ref{eq:indeq}) (see
also Ref.~\cite{AkhDvo24}) for the magnetic field $\mathbf{B}$,
which accounts for the inhomogeneity of the pseudoscalar field, $\nabla\varphi\neq0$.
Equation~(\ref{eq:indeq}) is completed with the Klein-Gordon Eq.~(\ref{eq:KGeq})
with the nonzero right hand side describing the interaction between
$\varphi$ and the electromagnetic field.

Then, we have developed the axion dynamo in a thin spherical layer.
Using the symmetry properties of the poloidal and toroidal magnetic
fields, as well as those of $\varphi$, and neglecting the radial
dependence of the fields, which is a standard dynamo approximation
(see, e.g., Ref.~\cite{Par55}), we have derived the full set of
the evolution equations. These equations have been rewritten in the
dimensionless variables in Eq.~(\ref{eq:ABPhieq}). We have used
the low mode approximation, accounting for two harmonics (see Ref.~\cite{NefSok10}),
in Eq.~(\ref{eq:ABPhiharm}) to reduce the general evolution equations
to the system of nonlinear ordinary differential equations. The details
are present in Appendix~\ref{sec:DIFFEQHARM}.

In Sec.~\ref{sec:APPL}, we have studied the application of our results
for the description of the magnetic field evolution inside a small
axion star embedded in solar plasma. For this purpose, we have considered
a dense axion spherical structure with $R\sim1\,\text{cm}$, which
was predicted in Ref.~\cite{Bra16}. The seed magnetic field has
been taken as $4\,\text{kG}$. We have considered the turbulent magnetic
diffusion coefficient corresponding to the observational value~\cite{ChaLitSak08}.

We have obtained that the magnetic field enters to the oscillations
regime. The typical frequency of magnetic field oscillations is $\sim10^{10}\,\text{Hz}$.
This frequency guarantees to validity of both the MHD approximation
and the causality condition. Such frequencies are covered by the modern
solar radio telescopes (see, e.g., Ref.~\cite{Sai12}). Thus, potentially
the related electromagnetic radiation can be observed.

When such small axionic objects decay, the confined energy of oscillating
magnetic fields is liberated as electromagnetic waves. Spatially localized
electromagnetic flashes with the frequency $f\lesssim160\,\text{MHz}$
were reported in Ref.~\cite{Mon20} to be a possible source of the
solar corona heating. This frequency is slightly below our prediction,
especially if we consider the greatest first peaks in the spectra
of $a_{1}$ in Figs.~\ref{fig:fig1c} and~\ref{fig:fig2c}.
The connection of the observational data in Ref.~\cite{Mon20} with
the annihilation of dark matter nuggets~\cite{Zhi03} was discussed
in Ref.~\cite{Ge20}. The review of solar radio emission caused by
the dark matter is given in Ref.~\cite{An23}. We suggest that small
size axion stars, which contain the internal oscillating magnetic
fields, described in the present work, can be a possible explanation
of flashes in the Sun observed in Ref.~\cite{Mon20}.

\section*{Acknowledgments}

I am thankful to D.~D.~Sokoloff for the useful discussion.

\appendix

\section{Derivation of the modified induction equation\label{sec:DERINDEQ}}

The axion electrodynamics results from the following Lagrangian~\cite{DvoSem20}:
\begin{equation}\label{eq:Lagr}
  \mathcal{L}=-\frac{1}{4}F_{\mu\nu}F^{\mu\nu}+\frac{1}{2}(\partial_{\mu}\varphi\partial^{\mu}\varphi-m^{2}\varphi^{2})-
  \frac{g_{a\gamma}\varphi}{4}F_{\mu\nu}\tilde{F}^{\mu\nu}-A^{\mu}J_{\mu},
\end{equation}
where $F_{\mu\nu}=(\mathbf{E},\mathbf{B})$ is the electromagnetic
field tensor, $\tilde{F}_{\mu\nu}=\tfrac{1}{2}\varepsilon_{\mu\nu\lambda\rho}F^{\lambda\rho}$
is the dual tensor, $A^{\mu}$ is the electromagnetic field potential,
$J^{\mu}=(\rho,\mathbf{J})$ is the external current. The modified
Maxwell equations, coming from Eq.~(\ref{eq:Lagr}), have the form~\cite{Dvo22},
\begin{align}
  (\nabla\times\mathbf{B}) & =
  \frac{\partial\mathbf{E}}{\partial t}+\mathbf{J}+g_{a\gamma}\mathbf{B}\frac{\partial\varphi}{\partial t}+
  g_{a\gamma}(\nabla\varphi\times\mathbf{E}),
  \label{eq:Maxeq1}
  \\
  (\nabla\times\mathbf{E}) & =-\frac{\partial\mathbf{B}}{\partial t},
  \label{eq:Maxeq2}
  \\
  (\nabla\cdot\mathbf{E}) & =-g_{a\gamma}(\mathbf{B}\cdot\nabla)\varphi+\rho,
  \label{eq:Maxeq3}
  \\
  (\nabla\cdot\mathbf{B}) & =0.
  \label{eq:Maxeq4}
\end{align}
We suppose that plasma is electroneutral, i.e. $\rho=0$ in Eq.~(\ref{eq:Maxeq3}).
Equation~(\ref{eq:Maxeq1}) should be completed by the Ohm's law
$\mathbf{J}=\eta^{-1}\mathbf{E}$, where we omit the advection term
since we study a slowly rotating axion star. Such a term was accounted
for in Ref.~\cite{Anz23}. Moreover, we neglect the displacement
current $\frac{\partial\mathbf{E}}{\partial t}$ with respect to the
Ohmic current in Eq.~(\ref{eq:Maxeq1}), which is a usual MHD approximation.

After these assumptions, Eq.~(\ref{eq:Maxeq1}) becomes algebraic
for the electric field. The electric field can be found in the form~\cite{Dvo22,AkhDvo24},
\begin{equation}\label{eq:Egen}
  \mathbf{E}=\eta(\nabla\times\mathbf{B})-g_{a\gamma}\eta\frac{\partial\varphi}{\partial t}\mathbf{B}-
  g_{a\gamma}\eta^{2}[\nabla\varphi\times(\nabla\times\mathbf{B})],
\end{equation}
which coincides with Eq.~(\ref{eq:E}). Note that we keep only the
terms linear in the coupling constant $g_{a\gamma}$ in Eq.~(\ref{eq:Egen}).

Based on Eqs.~(\ref{eq:Maxeq2}) and~(\ref{eq:Egen}), we obtain
the modified induction equation for the magnetic field~\cite{Dvo22,AkhDvo24},
\begin{equation}\label{eq:indeqgen}
  \frac{\partial\mathbf{B}}{\partial t}=\nabla\times
  \left[
    g_{a\gamma}\eta^{2}\nabla\varphi\times(\nabla\times\mathbf{B})+
    g_{a\gamma}\eta\frac{\partial\varphi}{\partial t}\mathbf{B}-\eta(\nabla\times\mathbf{B})
  \right],
\end{equation}
which is represented in Eq.~(\ref{eq:indeq}). Note that $\mathbf{B}$
in Eq.~(\ref{eq:indeqgen}) automatically satisfies Eq.~(\ref{eq:Maxeq4}),
i.e. it is divergenceless.

\section{Differential equations for the harmonics\label{sec:DIFFEQHARM}}

In this appendix, we derive the system of ordinary differential equations
for the evolution of the coefficients $a_{1,2}$, $b_{1,2}$, and
$\phi_{1,2}$.

For this purpose, we insert Eq.~(\ref{eq:ABPhiharm}) into Eq.~(\ref{eq:ABPhieq}).
Then, we multiply each equation by the corresponding function $\sin n\theta$,
where $n=1,\dots4$, and integrate the result, $\tfrac{2}{\pi}\smallint_{0}^{\pi}\dots\mathrm{d}\theta$,
taking into account the orthonormality condition,
\begin{equation}
  \frac{2}{\pi}\int_{0}^{\pi}\sin(n\theta)\sin(l\theta)\mathrm{d}\theta=\delta_{nl},
\end{equation}
where $n,l=1,\dots4$.

Finally, we obtain the following nonlinear differential equations:
\begin{align}\label{eq:abpsieq}
  \dot{a}_{1}= & -2(a_{1}+a_{2})+\frac{2}{\pi}
  \bigg[
    -\frac{64}{15}b_{1}\phi_{1}-\frac{1024}{63}b_{2}\phi_{2}+\frac{16}{15}b_{1}\psi_{1}+\frac{64}{63}b_{2}\psi_{2}-
    \frac{32}{105}b_{1}\psi_{2}
    \notag
    \\
    & -
    \frac{32}{105}b_{2}\psi_{1} +
    \frac{512}{105}b_{1}\phi_{2}+\frac{128}{105}b_{2}\phi_{1}
  \bigg],
  \nonumber
  \displaybreak[2]
  \\
  \dot{a}_{2}= & -12a_{2}+\frac{2}{\pi}
  \bigg[
    -\frac{64}{35}b_{1}\phi_{1}-\frac{20992}{3465}b_{2}\phi_{2}+\frac{16}{21}b_{1}\psi_{1}+\frac{64}{165}b_{2}\psi_{2}+
    \frac{32}{45}b_{1}\psi_{2}
    \notag
    \\
    & +
    \frac{32}{45}b_{2}\psi_{1}-\frac{256}{105}b_{1}\phi_{2}-\frac{2176}{315}b_{2}\phi_{1}
  \bigg],
  \nonumber
  \displaybreak[2]
  \\
  \dot{b}_{1}= & -2(3b_{1}+2b_{2})+\frac{2}{\pi}
  \bigg[
    \frac{592}{105}a_{2}\psi_{1}+\frac{608}{315}a_{2}\psi_{2}+\frac{288}{7}a_{2}\phi_{2}+\frac{352}{105}a_{1}\psi_{2}+
    \frac{304}{105}a_{2}\phi_{1}
    \notag
    \\
    & +
    \frac{112}{15}a_{1}\phi_{1}-\frac{608}{105}a_{1}\phi_{2}+\frac{16}{15}a_{1}\psi_{1}
  \bigg],
  \nonumber
  \\
  \dot{b}_{2}= & -20b_{2}+\frac{2}{\pi}
  \bigg[
    \frac{1888}{315}a_{2}\psi_{1}+\frac{11584}{3465}a_{2}\psi_{2}+\frac{294592}{3465}a_{2}\phi_{2}+\frac{64}{63}a_{1}\psi_{2}+
    \frac{288}{7}a_{2}\phi_{1}
    \notag
    \\
    & -
    \frac{608}{105}a_{1}\phi_{1}+\frac{1984}{63}a_{1}\phi_{2}-\frac{416}{105}a_{1}\psi_{1}
  \bigg],
  \nonumber
  \\
  \dot{\psi}_{1}= & -(\mu^{2}+2\kappa^{2})\phi_{1}+\frac{2}{\pi}
  \left[
    \frac{6688}{315}a_{2}b_{2}-\frac{32}{15}a_{1}b_{2}+\frac{2096}{105}a_{2}b_{1}+\frac{112}{15}a_{1}b_{1}
  \right],
  \nonumber
  \\
  \dot{\psi}_{2}= & -(\mu^{2}+12\kappa^{2})\phi_{2}+4\kappa^{2}\phi_{1}+\frac{2}{\pi}
  \left[
    \frac{65216}{3465}a_{2}b_{2}+\frac{2752}{315}a_{1}b_{2}+\frac{5408}{315}a_{2}b_{1}+\frac{544}{105}a_{1}b_{1}
  \right],
\end{align}
where $\psi_{1,2}=\dot{\phi}_{12}$ and a dot means the $\tau$ derivative.

\end{document}